\documentclass[fleqn,twoside]{article}
\usepackage{espcrc2}
\usepackage{epsfig}

\usepackage{graphicx}
\usepackage[figuresright]{rotating}

\newcommand{\AmS}{{\protect\the\textfont2
  A\kern-.1667em\lower.5ex\hbox{M}\kern-.125emS}}

\title{Density matrix renormalization group approach 
to a two-dimensional bosonic model}

\author{Takanori Sugihara\address[MCSD]{RIKEN BNL Research Center, 
        Brookhaven National Laboratory, 
        Upton, NY 11973, U.S.A.}}
       
\begin{document}

\begin{abstract}
Density matrix renormalization group (DMRG) is applied to 
a (1+1)-dimensional $\lambda\phi^4$ model 
to study spontaneous breakdown of discrete $Z_2$ symmetry numerically. 
We obtain the critical coupling 
$(\lambda/\mu^2)_{\rm c}=59.89\pm 0.01$ 
and the critical exponent $\beta=0.1264\pm 0.0073$, which are 
consistent with the Monte Carlo and the exact results, respectively. 
The results are based on extrapolation to the continuum limit 
with lattice sizes $L=250,500$, and $1000$. 
We show that the lattice size $L=500$ is sufficiently 
close to the the limit $L\to\infty$ \cite{Sugihara:2004qr}. 
\vspace{1pc}
\end{abstract}

\maketitle

\section{Introduction}

Hamiltonian diagonalization is a useful method 
for nonperturbative analysis of many-body 
quantum systems. 
If Hamiltonian is diagonalized, the system can be analyzed 
nonperturbatively at the amplitude level
\cite{Harada:1993va,Sugihara:1997xh,Sugihara:2001ch}. 
However, in general quantum field theories, 
the method does not work without reducing degrees of freedom 
because the dimension of Hamiltonian increases exponentially 
as the system size becomes large. 
We need to find a way to create a small number of optimum 
basis states \cite{Wilson:1974mb,Sugihara:2001ci}. 
S.~White proposed a powerfull method 
called density matrix renormalization group (DMRG) \cite{white,lec}. 
In DMRG, calculation accuracy of target states can be 
controlled systematically using density matrices. 
White calculated very accurately energy and wavefunctions of 
Heisenberg chains composed of more than 100 sites 
using a standard workstation. 
The calculation reproduced the exact value of ground-state energy 
in five digits or higher. 
DMRG has been applied to various one-dimensional models, 
such as Kondo, Hubbard, and $t$-$J$ chain models, 
and achieved great success. In many cases, 
DMRG can give more accurate results than quantum Monte Carlo. 
A two-dimensional Hubbard model has also been studied with DMRG 
in both real- and momentum-space representation \cite{xiang,xls}. 
DMRG works well on small two-dimensional lattices and new 
techniques have been proposed for solving larger lattices \cite{tpva}. 
DMRG has also been extended to finite-temperature 
chain models using the transfer-matrix technique \cite{st}. 
In particle physics,  the massive Schwinger model has been 
studied using DMRG to confirm the well-known Coleman's picture 
of `half-asymptotic' particles 
at a background field $\theta=\pi$ \cite{Byrnes:2002nv}.
It would be interesting to seek a possibility of applying 
the method to QCD in order to study color confinement 
and spontaneous chiral symmetry breaking 
based on QCD vacuum wavefunctions. 

DMRG was originally proposed as a method for spin and fermion chain models. 
In fermionic lattice models, the number of particles 
contained in each site is limited because of the Pauli principle. 
On the other hand, in bosonic lattice models, each site can contain 
infinite number of particles in principle. 
It is not evident whether Hilbert space can be described 
appropriately with a finite set of basis states in bosonic models.
This point becomes crucial when DMRG is applied to 
gauge theories because gauge particles are bosons. 
Before working in lattice gauge theories 
like Kogut-Susskind Hamiltonian \cite{ks}, 
we need to test DMRG in a simple bosonic model 
and recognize how many basis states are necessary for each site 
to reproduce accurate results. 
In this work, we apply DMRG to a $\lambda\phi^4$ model 
with (1+1) space-time dimensions. 
We construct a Hamiltonian model on a spatial lattice.  
The model has spontaneous breakdown of discrete $Z_2$ symmetry 
and the exact values of the critical exponents are known. 
We are going to justify the relevance of DMRG truncation 
of Hilbert space in the bosonic model 
by comparing our numerical results with the Monte Carlo 
and the exact results \cite{Loinaz:1997az,onsager}. 
Our calculations are done with lattice sizes $L=250, 500$ and $1000$. 
$L=1000$ is about twice of the latest Monte Carlo one \cite{Loinaz:1997az}.

\section{DMRG in the $\lambda\phi^4_{1+1}$ model}
\begin{table*}[htb]
\caption{Various results for the critical coupling constant 
$(\lambda/\mu^2)_{\rm c}$ are listed.} 
\label{ccc}
\begin{tabular}{lcc}
\hline
Method & Result & Reference\\
\hline
DMRG                         & $59.89\pm 0.01$ & This work\\
Monte Carlo                  & $61.56_{-0.24}^{+0.48}$ & \cite{Loinaz:1997az}\\
Gaussian effective potential & $61.266$ & \cite{chang}\\
Gaussian effective potential & $61.632$ & \cite{Hauser:mb}\\
Connected Green function     & $58.704$ & \cite{Hauser:mb}\\
Coupled cluster expansion    & $22.8<(\lambda/\mu^2)_{\rm c}<51.6$ 
& \cite{Funke:wb}\\
Non-Gaussian variational     & $41.28$ & \cite{Polley:wf}\\
Discretized light cone       & $43.896$, $33.000$ &
 \cite{Harindranath:db}\\
Discretized light cone       & $42.948$, $46.26$ &
 \cite{Sugihara:1997xh}\\ \hline
\end{tabular}
\end{table*}
In the $\lambda\phi^4_{1+1}$ model, 
we divide a Hamiltonian
\begin{equation}
  \tilde{H} = \sum_{n=1}^L h_n + \sum_{n=1}^{L-1} h_{n,n+1}, 
  \label{ham}
\end{equation}
into two parts
\begin{eqnarray*}
  h_n &=& \frac{1}{2}\pi^2_n + \frac{\tilde{\mu}_0^2}{2}\phi_n^2
  +\frac{\tilde{\lambda}}{4!}\phi_n^4, 
  \nonumber
  \\
  h_{n,n+1} &=& \frac{1}{2}(\phi_n-\phi_{n+1})^2. 
\end{eqnarray*}
The field operator $\pi_n\equiv a\dot{\phi}_n$ is 
conjugate to $\phi_n$, $[\phi_m(t),\pi_n(t)] = i\delta_{mn}$. 
The derivative has been replaced with a naive difference. 
(Errors associated with the difference can be 
discussed if necessary \cite{Sugihara:2003mh}.) 
We rewrite Hamiltonian (\ref{ham}) using
creation and annihilation operators $a_n^\dagger$ and $a_n$. 
\begin{equation}
  \phi_n = \frac{1}{\sqrt{2}}
    \left(a_n^\dagger + a_n\right), \quad
  \pi_n = \frac{i}{\sqrt{2}}
    \left(a_n^\dagger - a_n\right), 
  \label{ca}
\end{equation}
where $[a_m,a_n^\dagger]=\delta_{mn}$ and $a_n|0\rangle=0$. 
Note that $a_n^\dagger$ and $a_n$ are not creation 
and annihilation operators in Fock representation. 
The index $n$ of the operators $a_n^\dagger$ and $a_n$ 
stands for the discretized spatial coordinate, not momentum. 
Real-space representation is better for our purpose 
because local interactions are useful for DMRG. 

The finite system algorithm of DMRG is applied 
to the Hamiltonian. 
A superblock Hamiltonian $H_{\rm S}$ 
is composed of two blocks and one site: 
\begin{equation}
  H_{\rm S} = \bar{H}_{\rm L} + h_{n-1,n} + h_n + h_{n,n+1}
  + \bar{H}_{\rm R}, 
\end{equation}
where $\bar{H}_{\rm L}$ and $\bar{H}_{\rm R}$ are 
effective Hamiltonian for the left and right blocks, respectively. 
$h_{n-1,n}$ ($h_{n,n+1}$) is an interaction 
between the left (right) block and the inserted $n$-th bare site. 
\begin{figure}[htb]
\epsfig{file=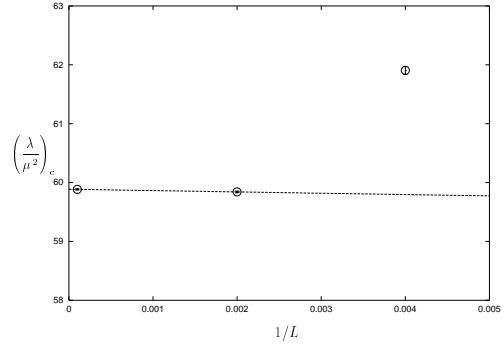,width=6.5cm}
\caption{$(\lambda/\mu^2)_{\rm c}$
is plotted as a function of $1/L$
for $L=250,500$, and $1000$. 
Extrapolation to the limit $L\to\infty$ gives 
$(\lambda/\mu^2)_{\rm c}=59.89\pm 0.01$. }
\label{cc_cntlim}
\end{figure}
\noindent
\begin{figure}[htb]
 \epsfig{file=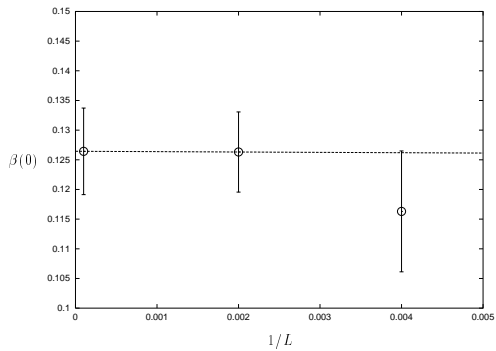,width=6.5cm}
\vspace{-0.5cm}
\caption{$\beta$ is plotted as a function of $1/L$ 
for $L=250,500$, and $1000$. 
Extrapolation to the limit $L\to\infty$ 
gives $\beta=0.1264\pm 0.0073$.}
\label{ce_cntlim}
\end{figure}
The target state is expanded as 
\begin{equation}
  |\Psi\rangle = \sum_{i=1}^M \sum_{j=1}^N \sum_{k=1}^M
  \Psi_{ijk}|i,j,k\rangle_n. 
\end{equation}
where the index $j$ is for the inserted bare site and 
$i$ and $k$ are for the renormalized blocks. 
The relevance of truncation with $M$ and $N$ 
is checked numerically by seeing convergence of 
energy and wavefunction. 
Parameter values $(M,N)=(10,10)$, which give good convergence, 
are used in all calculations. 
In Fig. \ref{cc_cntlim} and \ref{ce_cntlim}, 
we extrapolate the results to the continuum limit and 
obtain the critical coupling 
$(\lambda/\mu^2)_{\rm c}=59.89\pm 0.01$ 
and the critical exponent $\beta=0.1264\pm 0.0073$, 
which are consistent with the Monte Carlo 
$(\lambda/\mu^2)_{\rm c}=61.56_{-0.24}^{+0.48}$ 
and the exact $\beta=0.125$ results, respectively.

\section{Conclusion}
We have determined the critical coupling constant 
$(\lambda/\mu^2)_{\rm c}$ and the critical exponent $\beta$ 
of the model by extrapolating the numerical results for finite 
but sufficiently large lattices to the continuum limit. 
DMRG truncation works well also in the bosonic model. 
The lattice with $L=500$ can give results sufficiently 
close to the limit $L\to\infty$. 

The numerical calculations were carried on RIKEN VPP and 
Yukawa Institute SX5.

\end{document}